\documentclass[a4paper,fleqn,usenatbib]{mnras}

\usepackage[T1]{fontenc}
\usepackage{ae,aecompl}

\usepackage{graphicx}	
\usepackage{amsmath}	
\usepackage{amssymb}	
\usepackage{url}
\usepackage{float}
\usepackage{mathrsfs}
\usepackage{mathtools}
\usepackage{siunitx}
\usepackage{bigints}
\usepackage{longtable,multirow}
\usepackage{tabularx}
\usepackage{footnote}
\usepackage{tablefootnote}
\usepackage{subfig}
\usepackage{placeins}
\usepackage{listings}
\usepackage{color,hyperref}
\usepackage{gensymb}
\usepackage{pdflscape}
\usepackage{verbatim}
\usepackage{textgreek}
\usepackage{newtxtext}                  
\usepackage[slantedGreek]{newtxmath} 

\newcommand{\lxp}{\emph{AstroSat}/LAXPC}

\newcommand{\astrosat}{\emph{AstroSat}}
\newcommand{\rxte}{\emph{RXTE}}
\newcommand{\exosat}{\emph{EXOSAT}}

\newcommand{\insight}{\emph{Insight-HXMT}}
\newcommand{\hakucho}{\emph{Hakucho}}
\newcommand{\sas}{\emph{SAS-3}}
\newcommand{\lmxb}{4U 1636\ensuremath{-}536}

\title[AstroSat observation of \lmxb]{Thermonuclear X-ray bursts from \lmxb~ observed with \astrosat}

\author[Pinaki Roy et al.]{Pinaki Roy,$^{1}$\thanks{E-mail: pinaki.roy1989@gmail.com} Aru Beri,$^{1,2}$\thanks{E-mail: a.beri@soton.ac.uk}
Sudip Bhattacharyya$^{3}$\\\\
$^{1}$Department of Physical Sciences, IISER Mohali, Punjab 140306, India\\
$^{2}$School of Physics and Astronomy, University of Southampton, Southampton, Hampshire, SO17 1BJ United Kingdom\\
$^3$Department of Astronomy and Astrophysics, Tata Institute of Fundamental Research, 1 Homi Bhabha Road,~Mumbai 400005, India
}

\date{Accepted XXX. Received YYY; in original form ZZZ}

\pubyear{2021}

\begin{document}
\label{firstpage}
\pagerange{\pageref{firstpage}--\pageref{lastpage}}
\maketitle

\begin{abstract}

We report results obtained from the study of 12 thermonuclear X-ray bursts in 6 \astrosat~observations of a neutron star X-ray binary and well-known X-ray burster, \lmxb.~Burst oscillations at $\sim$\,581 Hz are observed with 4--5${\sigma}$ confidence in three of these X-ray bursts.~The rising phase burst oscillations show a decreasing trend of the fractional rms amplitude at $3{\sigma}$ confidence, by far the strongest evidence of thermonuclear flame spreading observed with~\astrosat.~During the initial 0.25 second of the rise a very high value ($34.0\pm6.7{\%}$) is observed.~The concave shape of the fractional amplitude profile provides a strong evidence of latitude-dependent flame speeds, possibly due to the effects of the Coriolis force.~We observe decay phase oscillations with amplitudes comparable to that observed during the rising phase, plausibly due to the combined effect of both surface modes as well as the cooling wake.~The Doppler shifts due to the rapid rotation of the neutron star might cause hard pulses to precede the soft pulses, resulting in a soft lag.~The distance to the source estimated using the PRE bursts is consistent with the known value of $\sim$~6~\rm{kpc}.

\end{abstract}




\section{Introduction}

Thermonuclear (Type~I) X-ray bursts are thought to be sudden eruptions in X-rays due to the unstable burning of hydrogen and helium on the surface of an accreting neutron star in a Low-Mass X-ray Binary (LMXB). Typically, they are characterized by a sharp increase in X-ray intensity and an exponential decay. The burst rise occurs in about 0.5--5~${\rm{s}}$ while the decline happens in around 10--100~$\rm{s}$ \citep{Lewin1977, Hoffman1978, Lewin1993, Galloway2008, Bhattacharyya2010}.~Fast-timing capability of the \emph{Rossi~X-ray Timing Explorer~(\emph{RXTE})} \citep{Jahoda06} led to the discovery of narrow but high-frequency features (mostly in the range: $\sim$\,300--600~\rm{Hz}) in the power spectrum of these X-ray bursts \citep[see, e.g.,][]{Strohmayer97a, Strohmayer97b}.~These features are termed as burst oscillations~(BOs), either found during their rising/decay phase or in both phases. Thermonuclear X-ray bursts have been observed in more than 100 LMXBs harboring a neutron star, and BOs have been found in quite a few sources \citep[see][and references therein]{Sudip2021}, and there has been a relatively quiet period after the termination of the \emph{RXTE} mission in 2011.~Searching for BOs requires a sensitive instrument, operating in a wide energy band and capable of providing ${\mu}{\rm{s}}$ time resolution.~After the launch of \emph{AstroSat} \citep{Singh2016} and \emph{Neutron Star Interior Composition Explorer}~(\emph{NICER}) \citep{Arzoumanian14}, the hunt for BOs began once again.\\

Although it is not straightforward to find BOs and constrain their properties, it is important to establish a complete picture of BOs as possible.~These are considered to result from the stellar rotation induced modulation of a brightness asymmetry so that pulsations are seen near the spin frequency \citep[see, e.g.,][]{Strohmayer1996, Watts2012}.~Several theories have been proposed to explain these features. Rising phase oscillations are understood due to the flame spreading from the ignition point of the bursts \citep[see, e.g.,][]{Strohmayer1997} while burst decay oscillations have been explained due to cooling wakes \citep[see, e.g.,][for details]{Mahmoodifar2016}.~However,~none of these explain all the observed BO properties \citep[see the review by][]{Watts2012}.~These features' frequency can evolve by $1\%$ (of the mean frequency) during a burst, generally from a lower initial value to an asymptotic value \citep{Strohmayer2006}. In some instances, a downward drift in frequency has also been noted \citep{Strohmayer1999b, Muno2000}.~Energy dependence study of BOs is essential to probe the origin of these features.~Attempts have been made in the past to understand phase lags of oscillations \citep[see, e.g.,][]{Muno2003, Artigue13}.~Several possibilities have been discussed in the literature that might be the cause for the soft lag of oscillations \citep[see, e.g.,][]{Cui_1998, Ford_1999, Sazonov_2001}.~Explanation for hard lags is a bit challenging though, as it is not clear if hard lags are due to Compton upscattering of photons which should decrease the fractional amplitude at higher energies, but this is in contrast to that observed \citep{Muno2003}.~It has also been recently found that different bursts from the same source show a diverse behavior, indicating no lag, hard lag, soft lag or mixed lag \citep{Chakraborty2017}.\\

Time-resolved spectroscopy during these X-ray bursts have been performed for several sources. The continuum spectra of Type~I X-ray bursts is often modelled using an absorbed blackbody, assuming that the entire surface of neutron star emits like a blackbody \citep[see, e.g.,][]{Swank1977,vanParadijs1978,Kuulkers2003}. In a conventional method, persistent emission prior to the X-ray burst (also referred to as preburst emission) is assumed to be constant (non-evolving) and subtracted as a background \citep[e.g.][]{Lewin1993,Bhattacharyya2010}. Recent studies have found limitations to this assumption, and persistent emission is seen to vary during these X-ray bursts \citep{Worpel2013,Worpel2015}.\\

There exists a particular class of X-ray bursts called Photospheric Radius Expansion (PRE) burst during which the flux approaches the local Eddington limit causing the photosphere to expand due to radiation pressure \citep{Tawara1984}.~A decrease in the blackbody temperature characterizes these bursts while the inferred blackbody radius simultaneously increases. All this happens at an approximately constant value of total flux \citep[see][for details]{Galloway2008}. When the photosphere returns to the neutron star surface, the temperature has the highest value against a very low blackbody radius value. This stage is called the touchdown \citep[see, e.g.,][]{Kuulkers2003}. PRE bursts can be distinguished from typical Type I X-Ray bursts by two maxima in the temperature profile and a burning area maximum corresponding to the temperature minimum \citep[see, e.g.,][]{Galloway2008,Bhattacharyya2010}. These bursts can serve as distance indicators \citep[see, e.g.,][]{Basinska1984}.\\

In this paper, we report X-ray bursts in \lmxb\ observed with \astrosat.~\lmxb\ is a neutron star LMXB with an orbital period of $\sim$\,3.8 hr \citep{vanParadijs1990} and a main-sequence companion star of mass 0.4--0.5 $M_\odot$ \citep{Giles2002,Casares2006}.~The distance to the source is $\sim$\,6 kpc \citep{Galloway2006}. The neutron star's mass is estimated to be 1.6–2.0 $M_\odot$ \citep{Casares2006}.~\lmxb\ is classified as an atoll source \citep{Schulz1989,vanderKlis1989} based on the track it traces in the color-color diagram (CCD) and the hardness-intensity diagram (HID) \citep{Belloni2007,Altamirano2008b} on a $\sim$\,40-day cycle \citep{Shih2005,Belloni2007}. The transition through CCD or HID is believed to arise from alterations in the mass accretion rate with the source moving from a hard state to a soft state as the accretion rate increases \citep{Bloser2000,Gierlinski2002a,Gierlinski2002b}. These two states have different predispositions to different varieties of thermonuclear X-ray bursts \citep[see][for details]{Hoffman1977,Galloway2006,Zhang2011}.
As an example, short-recurrence bursts (burst-doublets and burst-triplets) tend to occur during the hard state \citep[e.g.,][]{Beri2019}, whereas PRE bursts and superbursts (very long X-ray bursts) as well as BOs are preferentially observed during the soft state \citep[e.g.,][]{Guver2012b,Galloway2017}.~For this source, time-resolved burst spectroscopy has been performed earlier using observations from \sas\ \citep{Hoffman1977}, \hakucho\ \citep{Ohashi1982}, \exosat\ \citep[e.g.,][]{Turner1984,Sztajno1985}, \rxte\ \citep[e.g.,][]{Strohmayer1999a,Galloway2006,Linares2009,Zhang2009}, \insight\ \citep{Chen2018} and \astrosat\ \citep{Beri2019}. Burst oscillations in this source have so far been reported with \rxte\ data \citep[e.g.,][]{Giles2002} but not with  \astrosat\ data. This paper attempts to fill that gap and we discuss our results in light of proposed theoretical models. Previously, burst oscillation has been reported for a Type~I burst in 4U 1728\ensuremath{-}34 using \astrosat\ observation \citep{VerdhanChauhan2017} demonstrating the capability of \lxp\ instrument for detecting millisecond variability in these sources.

\section{Observations and Data Reduction}

\subsection{\lxp}
\label{sec-obslxp}
AstroSat is India's first dedicated multi-wavelength astronomy satellite, launched on September 28, 2015, into a 650 km orbit inclined at an angle of $6\degree$ to the equator by a PSLV C-30 rocket from Satish Dhawan Space Centre (SDSC), SHAR. Its orbital period is 97.6 min (5856 sec). It has five science payloads which cover from X-ray to UV wavelengths \citep{Agrawal2006,Paul2013,Singh2016}.\\

Large Area X-ray Proportional Counter (\textsc{LAXPC}) unit consisting of three nominally identical mutually independent detectors labeled as \textsc{LAXPC10}, \textsc{LAXPC20}, and \textsc{LAXPC30}, work in the energy range of 3-80 keV and has an effective collecting area of 6000 cm$^2$ in the 5--20 keV band. The \textsc{LAXPC} detectors have a collimator with a field of view of $0.9\degree\times0.9\degree$. Each detector has 60 anode cells, arranged in 5 layers of 12 cells each, producing 7 anode outputs (2 each from layers 1 \& 2, and 1 each from layers 3, 4 \& 5) designated A1--A7. The detectors have a time resolution of 10 ${\mu}{\rm{s}}$ and a dead-time of approximately 43 ${\mu}{\rm{s}}$. The energy resolution for \textsc{LAXPC10}, \textsc{LAXPC20} and \textsc{LAXPC30} at 30 keV are 15\%, 16\% and 10\% respectively \citep[see \S 8 of][]{Antia2017}.\\

It is found that the energy resolution of \textsc{LAXPC10} is steady while that in \textsc{LAXPC20} it is degraded from 12 to 16\% at 30 keV. On the other hand, the energy resolution of \textsc{LAXPC30} improved from 11 to 10\% \citep[see][for details]{Antia2021}. In \cite{Beri2019}, observations made in February 2016 (just after the launch) were used. Therefore, quoted values of energy resolution are different to those mentioned in this work.\\

\textsc{LAXPC} data are collected in two different modes: Broad Band Counting (BBC) and Event Analysis mode (EA). The EA mode data contain information about the time, anodeID, and Pulse Height (PHA) of each event, which is why we use this mode of data for timing and spectral analysis. We use the \textsc{LAXPC} software (\textit{laxpcsoftv3.0}) \citep{Antia2017} to generate the total spectra, the background spectra and the response files. Since the software applies a suitable dead-time correction to the light curve and spectrum, we do not perform this correction separately.~Table \ref{list-obs} gives the log of observations.~In this work, we do not include the observation made between 15 and 16 February 2016, reported by \cite{Beri2019}, as BOs were not detected during the seven X-ray bursts observed.

\begin{table*}
\caption{\lxp\ observation details of \lmxb. For convenience, we also list which detectors are used in this study for timing (`T') and spectral (`S') analysis of an observation. `X' indicates `omitted'. (see \S\,\ref{sec-obslxp})}
\begin{tabular}{ccccccccc}
\hline
Observation ID & Observation & Observation & Exposure & Bursts & Burst labels & \multicolumn{3}{c}{Used for Analysis}\\
& labels & date & (\rm{ks}) & & & LXP10 & LXP20 & LXP30\\
\hline
G05\_195T01\_9000000530 & Obs~1 & 2016-07-02 & 39 & 2 & B1, B2 & S, T & S, T & T\\
G05\_195T01\_9000000598 & Obs~2 & 2016-08-13 & 38 & 1 & B3 & S, T & S, T & T\\
G06\_104T01\_9000001060 & Obs~3 & 2017-02-28 & 45 & 2 & B4, B5 & S, T & S, T & T\\
G07\_040T01\_9000001326 & Obs~4 & 2017-06-21 & 22 & 3 & B6, B7, B8 & S, T & S, T & T\\
G08\_033T01\_9000002084 & Obs~5 & 2018-05-09 & 11 & 2 & B9, B10 & S, T & S, T & X\\
G08\_033T01\_9000002278 & Obs~6 & 2018-08-06 & 11 & 2 & B11, B12 & X & S, T & X\\
\hline
\end{tabular}
\label{list-obs}
\end{table*}

\section{Analysis \& Results}

\begin{figure}
\centering
\includegraphics[width=\columnwidth]{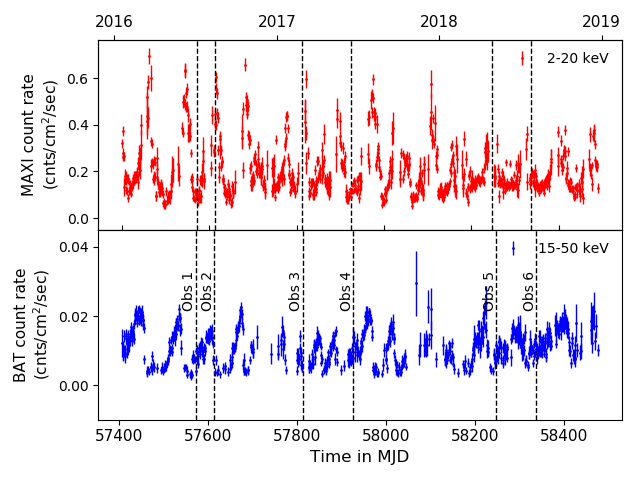}
\caption{\textit{MAXI}-\textsc{GSC} and \textit{Swift}-\textsc{BAT} light curves of \lmxb\ in 2--20 keV band and 15--50 keV band respectively with vertical dash lines indicating the \astrosat\ observations used in this work. A signal-to-noise ratio of 3 is used as a filter for both the light curves. (see \S\,\ref{sec-batlc})}
\label{batlc}
\end{figure}

\subsection{\emph{MAXI}-\textsc{GSC} and \emph{Swift}-\textsc{BAT} Light Curves of \lmxb }
\label{sec-batlc}
The \textit{Swift}\,/\,Burst Alert Telescope (\textsc{BAT}) is a hard X-ray transient monitor which observes $\sim$\,88\% of the sky each day with a detection sensitivity of 5.3 mCrab and a time resolution of 64 seconds, providing almost real-time coverage of the X-ray sky in the energy range 15-50 keV \citep{Krimm2013}. The Gas Slit Camera (GSC: \cite{Mihara2011}) onboard the Monitor of All-sky X-ray Image (MAXI: \cite{Matsuoka2009}) covers $\sim$\,85\% of the sky per 92-minute orbital period and $\sim$\,95\% per day with a detection sensitivity of 15 mCrab in the 2–20 keV band in a daily scan \citep{Sugizaki2011}. ~We use the \emph{MAXI} and \emph{Swift}-\textsc{BAT} light curves of \lmxb~to determine this source's spectral state (see Figure~\ref{batlc}). The two light curves of \lmxb\ are binned with a binsize of 1~day. A signal-to-noise ratio of 3 is used as a filter for both the light curves.~The peak of the \emph{MAXI} light curve (high value of flux in 2-20 keV band) indicates a soft spectral state, whereas, the peak of the \textsc{BAT} light curve (high value of flux in 15-50 keV band) suggests a hard spectral state of a source.

\subsection{Color-Color Diagram}
\label{sec-ccd}
Atoll sources mainly exhibit three tracks in the color-color diagram, namely, the extreme island state (EIS), the island state (IS) and the banana branch (BB). EIS and IS are spectrally harder states with lower X-ray luminosity whereas BB is the spectrally softer state with higher X-ray luminosity \citep{Altamirano2008b}.\\

\begin{figure}
\centering
\includegraphics[width=0.95\columnwidth]{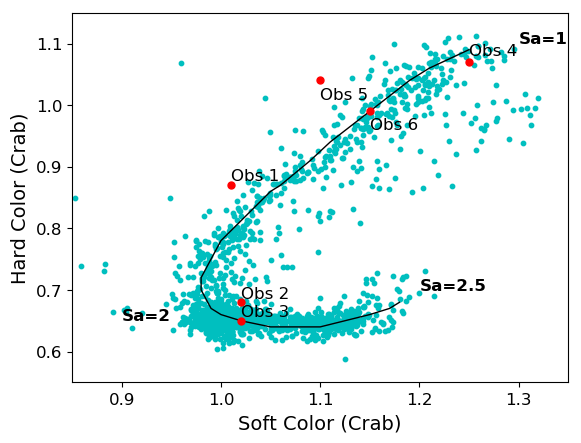}
\caption{Color-color diagram of \lmxb\ with positions of the six observations studied in this paper against the \rxte\ observations \citep[adapted from][]{Altamirano2008a} shown as cyan dots. The thin solid curve, $S_a$, parametrizes the position of the source on the diagram. (see \S\,\ref{sec-ccd})}
\label{color}
\end{figure}

To ascertain the spectral state of the 6 \textsc{LAXPC} observations, we plot the color-color diagram (Figure~\ref{color}) of all the \rxte\ observations of \lmxb\ as given in \citep{Altamirano2008a} and the location of the observations studied in this paper. For the \rxte\ observations, hard and soft colors are taken as the 9.7-16.0/6.0-9.7 keV and 3.5-6.0/2.0-3.5 keV count rate ratios, respectively.~The colors are normalized by the corresponding Crab Nebula color values that are closest in time to correct for the gain changes as well as for the differences in the effective area between different proportional counters. For the \astrosat\ observations, only \textsc{LAXPC20} is used to calculate the color values. The effective area curves for \emph{RXTE}-\textsc{PCA} and \textsc{LAXPC} are similar in the 3--25 keV range (refer to Figure~1 of \cite{Paul2013}) and we find that the Crab-normalized color values for \textsc{LAXPC20} are consistent with those reported with \rxte.~Therefore,~the definition of hard and soft colors is retained as that of \rxte. \\

The location of the source on the diagram is customarily parametrized by the length of the curve, $S_a$. It is normalized so that $S_a=1$ at the top-right end and $S_a=2$ at the bottom-left end of the diagram \citep[see, e.g.][]{Mendez1999}. $S_a$ is believed to represent the mass accretion rate \citep{Hasinger1989}.~From Figure~\ref{color}, it is clear that Obs~1,~Obs~4,~Obs~5 and ~Obs~6 belong to the hard spectral state while Obs~2 and Obs~3 belong to the soft state of the source.

\subsection{Timing Analysis \& Results}
\subsubsection{Energy-Resolved Burst Profile}
\label{sec-enlc}
\begin{figure}
\centering
\includegraphics[width=\columnwidth]{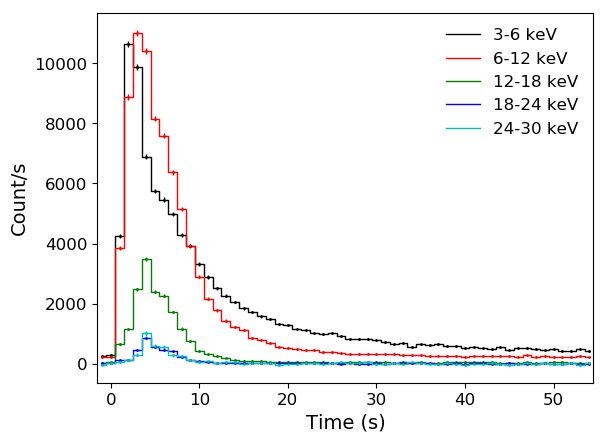}
\caption{(\textsc{LAXPC10}+\textsc{LAXPC20}+\textsc{LAXPC30}) background-corrected light curve of burst B4. Binsize is 1~\rm{s}. Count-rates in 24--30~\rm{keV} band are multiplied by a factor of 5. This figure shows that the X-ray burst is detected up to 30 keV. (see \S\,\ref{sec-enlc})}
\label{EN-lc}
\end{figure}

We find 12 thermonuclear X-ray bursts, and their energy-resolved profiles indicate that all bursts are significantly detected up to 25~\rm{keV}. To illustrate this, we show here the burst profile of one X-ray burst~(see Figure~\ref{EN-lc}). We use five narrow energy bands, viz. 3--6~{\rm{keV}}, 6--12~\rm{keV}, 12--18~\rm{keV}, 18--24~\rm{keV} and 24--30~\rm{keV}. The light curves are created using events from all layers of \textsc{LAXPC}. For the light-curve in the 24--30~\rm{keV} band, the count-rates are multiplied by a factor of 5 for visual clarity. 

\subsubsection{Power Spectrum: Burst Oscillations}
\label{sec-ps-bo}
We perform a search for <1024~{\rm{Hz}} (Nyquist frequency) oscillations along the entire duration of each of the 12 bursts. Events from all three \textsc{LAXPC} detectors are taken into account, however, \textsc{LAXPC30} is not used for timing analysis of the two 2018 observations (Obs~5 and Obs~6) owing to its defunct status. The events (photon arrival times) in the 3--25~\rm{keV} energy band are sampled at the Nyquist rate of 2048~\rm{Hz}. We perform fast Fourier transform (FFT) of successive 1~{\rm{s}} segment (shifting the 1 s time window) of the input barycentre-corrected merged event file corresponding to the burst time interval. The FFT scan is repeated with the start time offset by 0.5~\rm{s}. A sharp signal at $\sim$\,581 Hz is clearly seen during three of the bursts in the Leahy-normalized \citep{Leahy1983} power spectrum. In such bursts, we examine the region that shows the signal at $\sim$\,581 Hz and attempt to maximize the measured power, $P_\text{m}$, by varying the start and end points of the segment in steps of 0.1 s and trying segment lengths of 1~\rm{s}, 2~\rm{s} and 3~{\rm{s}} within a time window of 4~{\rm{s}} (30+20+10=60 overlapping segments). We also check two more energy bands: 3--8~{\rm{keV}} and 8--25~{\rm{keV}}. The number of trials is thus, $60\times3=180$. The single-trial chance probability i.e., the probability of obtaining $P_\text{m}$ solely due to noise, is then given by the survival function, $e^{-P_\text{m}/2}$, where $P_\text{m}$ is now the maximized power obtained through the trials. So, the significance is $x=e^{-P_\text{m}/2}\times180$, and the confidence level would be $X\sigma$, where $X=\sqrt{2}\,\text{erf}^{-1}(1-x)$. The rms fractional amplitude is given by $A=[N_\gamma/(N_\gamma-N_\text{bkg})]\sqrt{P_\text{s}/N_\gamma}$ \citep[see, e.g.,][]{Watts2012}. $P_\text{s}$ is the signal power which can be approximated from $P_\text{m}$ using Table 1 (with $k=1$) of \cite{Bilous2019} if $P_\text{m}\gtrsim 2(k+\sqrt{k})$. We use the median value $P_s|P_m$, so that $P_s=P_m+1$ and $2\sqrt{P_m+1}$ for the uncertainty on $P_s$ \citep[see \S\,4.5 of][]{Bilous2019}. $N_\gamma$ is the total number of photons, and $N_\text{bkg}$ is the number of background photons (estimated from the interval prior to the burst for the same duration and energy range as the FFT window).~The results are summarized in Table~\ref{BO-summary}. The power spectra showing $\sim$\,581~{\rm{Hz}} are included in Appendix (see Figure~\ref{A2-B3-B4-B9-BO} in Appendix A).

\begin{table}
\caption{\lmxb\ burst oscillations as observed with \lxp. The given values are for the power spectra shown in Figure~\ref{A2-B3-B4-B9-BO}. \textsc{LAXPC30} was unavailable during burst B9. For details, see \S\,\ref{sec-ps-bo}.}
\setlength{\tabcolsep}{3pt}
\renewcommand{\arraystretch}{1.4}
\begin{tabular}{cccccc}
\hline
Burst & Oscillation & Phase & Chance & Confidence & Fractional\\
& frequency & & probability\footnotemark[1] & level & amplitude\\
\hline
B3 & 580--581 Hz & decay & $1.38\times10^{-9}$ & $5.2\sigma$ & $6.8\pm1.0$\%\\
B4 & 580--581 Hz & decay & $4.85\times10^{-7}$ & $3.9\sigma$ & $4.6\pm0.8$\%\\
B9 & 580--581 Hz & rise & $6.51\times10^{-9}$ & $4.9\sigma$ & $13.8\pm2.2$\%\\
\hline
\end{tabular}
\footnotemark[1]{single trial}
\label{BO-summary}
\end{table}

\subsubsection{Energy \& Time dependence of BOs}
\label{sec-entimebo}
To study the energy dependence of $\sim$\,581 Hz oscillations, we model the oscillations with the function $A+B\sin{2\pi\nu t}$ (e.g., see Figure~\ref{A3-B9-BO-profile} in Appendix A). Here, $B/A$ gives the half-fractional amplitude. The rms fractional amplitude is given by $B/(A\sqrt{2})$, which is comparable to that obtained from the power spectrum. For every time-window that shows oscillation, we determine the best-fit values of A and B, for the frequency which maximizes the fractional amplitude. To check if the oscillation is detected in the chosen window and chosen energy bin, the F-test approach is adopted \citep[see][for details]{Chakraborty2014}.\\

\begin{figure}
\centering
\includegraphics[width=0.95\columnwidth]{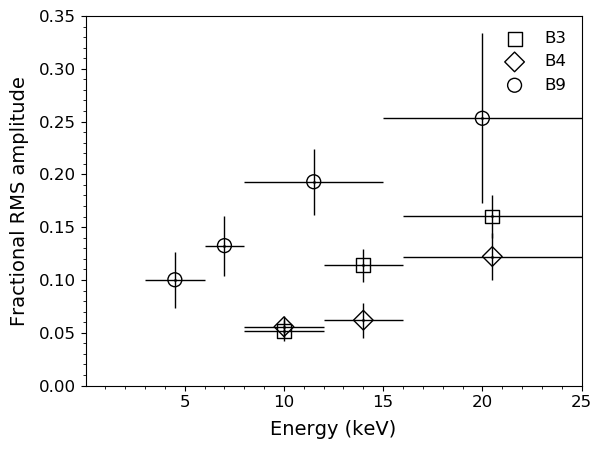}
\caption{Energy dependence of rms fractional amplitude in a 2 s window during the decay phase of bursts~B3 and B4 and in a 1 s window during the rising phase of burst B9. The vertical bars represent $1\sigma$ errors on amplitudes. The horizontal bars show the energy range. (see \S\,\ref{sec-entimebo})}
\label{Amplitude}
\end{figure}

For bursts~B3 and B4, the oscillations are mainly seen in the 8--25~\rm{keV} band, and sparsely in the 3--8~\rm{keV} band, although a similar number of counts are registered in either band. The variation in rms amplitudes within the 8–25 keV band shows an increasing trend with energy as shown in Figure~\ref{Amplitude}. For B3 and B4, 8--12 keV, 12--16 keV and 16--25 keV bands are used. In the case of burst B9, oscillations are detected in the 3--8~\rm{keV} band but with strength much less than that in the corresponding 8--25~\rm{keV} band. An increasing trend in rms amplitudes can be seen from Figure~\ref{Amplitude} wherein 3--6 keV, 6--8 keV, 8--15 keV and 15--25 keV bands are used for B9. The linear trends have slopes of 0.010, 0.007 and 0.010 keV$^{-1}$ for B3, B4 and B9 respectively.~Such an increasing trend in the rms amplitude of burst oscillations with energy is typical in non-pulsing systems \citep[see, e.g.,][]{Muno2003}.\\

\begin{figure}
\includegraphics[width=0.95\columnwidth]{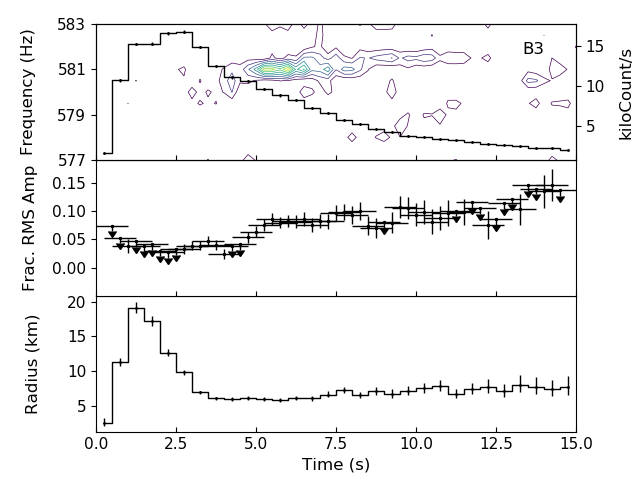}
\includegraphics[width=0.95\columnwidth]{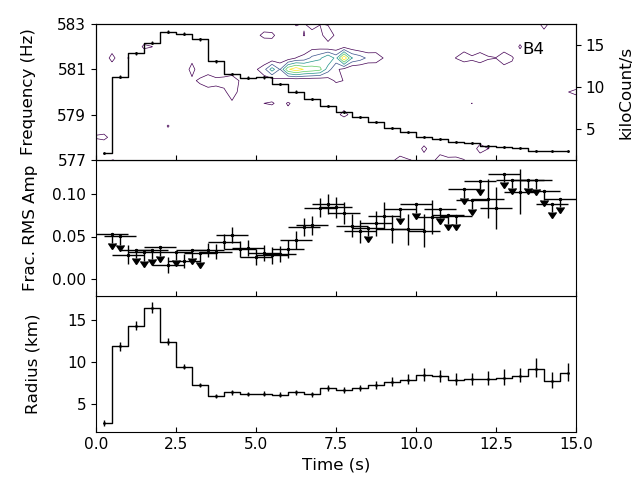}
\includegraphics[width=0.95\columnwidth]{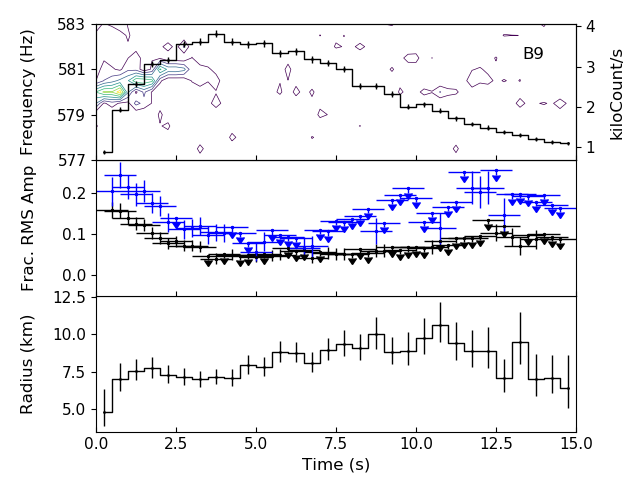}
\caption{The three plots starting from top are for the three bursts B3,~B4 and B9,~respectively. DPS and fractional rms amplitudes are drawn in 8--25 keV for bursts B3 and B4, and in 3--25 keV for B9. Amplitudes in 8--25 keV are also shown (in blue) for B9. The contours show Leahy normalized powers starting at 5 (\textit{outer contour}) and increasing in steps of 5. The down arrow marks the upper limit when oscillation is not detected. Refer \S\,\ref{sec-entimebo} for details.}
\label{time-evol-ampl}
\end{figure}

In Figure~\ref{time-evol-ampl}, we compare the time evolution of fractional amplitude to that of apparent blackbody radius for the decay phase (bursts B3 and B4) and the rising phase (burst B9). The first panel shows the power contours of the dynamic power spectrum of burst rise oscillations superposed on the 0.5~{\rm{s}}\,--\,binned burst profile. Here we use 2~{\rm{s}} windows with new windows starting at 0.25~{\rm{s}} intervals. The second panel shows the time evolution of fractional amplitude in 1~{\rm{s}} windows; new windows starting at 0.25~{\rm{s}} intervals. In case of no detection of oscillation, an upper limit is marked. A range of 3 Hz around the BO frequency is used to search for upper limits. The third panel shows the evolution of the apparent blackbody radius derived from time-resolved spectroscopy.\\

\begin{figure}
\centering
\includegraphics[width=0.95\columnwidth]{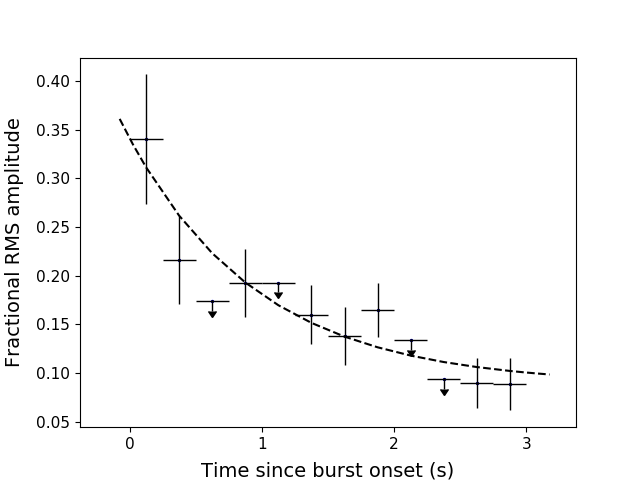}
\caption{Time evolution of fractional amplitude in 3--25 keV in 0.25~{\rm{s}} windows during the rise of burst B9. The vertical bars represent $1\sigma$ errors on amplitudes. The dashed curve shows the best-fit empirical model: $A=a-bc(1-e^{-t/c})$ \citep{Chakraborty2014} where $t$ is the time variable, and $a>0$, $b>0$, $c$ are the curve parameters. A positive value of $c$ indicates a concave profile. The best-fit value of c ($=0.99\pm0.52$) implies latitude-dependent flame speed, possibly due to the effects of the Coriolis force on flame spreading. (see \S\,\ref{sec-entimebo})}
\label{time-evol-ampl-rise}
\end{figure}

Next, we study the evolution of the fractional rms amplitude in 3--25 keV during the rise of burst B9. For this, we use 0.25 s binsize for amplitudes since close to the burst onset when the amplitude evolves rapidly, a small time bin is necessary to track the amplitude evolution \citep{Bhattacharyya2007}. Rapidly evolving amplitude during the rising phase also implies that the amplitude in a small time bin can be quite high compared to that in a large time bin which averages out the variation. We fit the rms amplitude curve (excluding the upper limits) with an empirical model: $A=a-bc(1-e^{-t/c})$ \citep{Chakraborty2014} where $t$ is the time variable, and $a>0$, $b>0$, $c$ are the curve parameters (see Figure~\ref{time-evol-ampl-rise}). The variable $c$ is called the convexity parameter, a positive value of which indicates a concave profile. The curve fitting is performed using \texttt{curve\_fit} from the SciPy package. The $\chi_{\nu}^2$ (d.o.f.) for the empirical model is 0.78 (5) and $c=0.99\pm0.52$. We also fit the profile with a constant model and obtain a $\chi_{\nu}^2$ (d.o.f.) of 4.27 (7). From the F-test between the two models we find that the empirical model is better than the constant model with a significance of $\approx 3\sigma$, thus emphasizing a decreasing trend. The value of the $c$-parameter implies a concave-shaped time evolution of rising phase rms amplitudes.

\begin{figure}
\centering
\includegraphics[width=0.95\columnwidth]{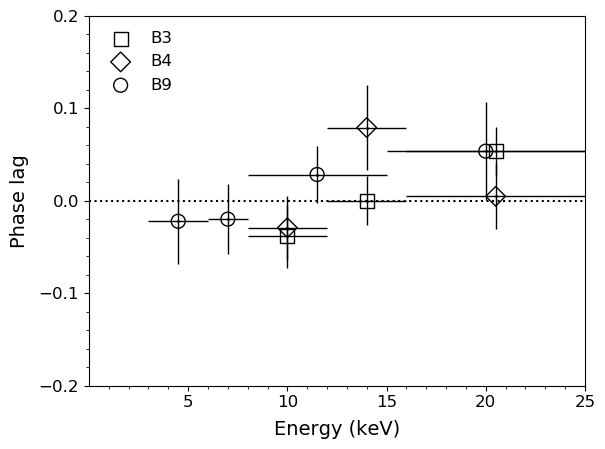}
\caption{Energy dependence of phase lag of oscillations in a 2 s window during the decay phase of bursts~B3 and B4 and in a 1 s window during the rising phase of burst B9. The vertical bars represent $1\sigma$ errors on phase lags. The horizontal bars show the energy range. A positive value of phase lag implies that the pulse in that energy band arrives earlier than the reference band. (see \S\,\ref{sec-plbo})}
\label{phase-lag}
\end{figure}

\subsubsection{Phase lag of BOs}
\label{sec-plbo}
In Figure~\ref{phase-lag}, we show the energy dependence of the phase lag of the oscillations in the three bursts for the same time windows and energy bands as in Figure~\ref{Amplitude}.~A positive slope, in our case, indicates a soft lag.~The phase is determined from the folded pulse profiles. The reference phase for calculating phase lag is taken from the folded pulse profile in 8--25 keV band for bursts B3 and B4 and 3--25 keV band for burst~B9. Taking respective first bands as reference also gives the same phase lag trends, however, the errors on the phase lag values gets progressively larger for higher bands.\\

In bursts B3 and B4, we find that the pulse in the lower band (8--12 keV) lags in phase behind the pulse in the upper band (12--16 keV) by 0.04 and 0.11 cycles, respectively, which translate to time delays of 70 and 190 ${\mu}{\rm{s}}$ respectively. In the burst B9 also, the pulse in the lower band (3–8 keV) lags in phase behind the pulse in the upper band (8--15 keV) by 0.05 cycles, equivalent to a time delay of 90 ${\mu}{\rm{s}}$.

\subsection{Time-resolved Spectroscopy during X-ray Bursts}
\label{sec-trs}
\subsubsection{Conventional Method}
\label{sec-trs-conv}
From energy-resolved light curves, we observe that the X-ray bursts are significantly detected up to 25 keV in \textsc{LAXPC10} and \textsc{LAXPC20} combined data. Therefore, we use the energy range of 3--25~keV for performing X-ray spectral fitting. Since in \textsc{LAXPC}, the soft and medium energy X-rays hardly reach the bottom layers of the detectors, time-resolved spectroscopy is done using single events from the top layer of each of the two detectors in order to minimize the background \citep[see][]{Beri2019,Sharma2020}. Data from \textsc{LAXPC30} are not included for spectral analysis due to its gain instability caused by gas leakage. For Obs 6, \textsc{LAXPC10} events are also excluded here due to gain loss in \textsc{LAXPC10} during this observation.~\\

To study the spectral evolution during X-ray bursts, the spectra during X-ray bursts are extracted for intervals of 0.5~{\rm{s}}. For each burst, we also extract the spectrum of 16~{\rm{s}} preceding the burst, which is subtracted for all the intervals in the burst as the underlying accretion emission and background. The spectra are grouped using \textsc{GRPPHA} to ensure a minimum of 25 counts per bin. The resulting burst spectra are fitted in \textsc{XSPEC version 12.10.1f} \citep{Arnaud1996} using the area normalized blackbody function, {`\texttt{BBODYRAD}'}, which has two variable parameters, viz. color temperature, $T_{\text{bb}}$, and normalization $K_{\text{bb}}=(R_{\text{bb}}/d_{10})^2$ where $R_{\text{bb}}$ is the apparent blackbody radius in {\rm{km}}, $d_{10}$ is the source distance in units of 10~{\rm{kpc}}. In order to model interstellar extinction, the {`\texttt{TBABS}'} component is used.~The photoionization cross sections and elemental abundances of {\texttt{TBABS}} are specified in \cite{Wilms2000}. The only variable parameter in {\texttt{TBABS}} is $N_H$, the Hydrogen column density, which is set to $0.25\times10^{22}$ cm$^{-2}$ \citep{Asai2000}. To account for cross-calibration between \textsc{LAXPC10} and \textsc{LAXPC20} (for Obs 1--5), a multiplicative constant component is included in the model. A systematic error of 1\% is added while performing the spectral fitting \citep{Antia2017}.\\

\begin{figure}
\centering
\includegraphics[width=\columnwidth]{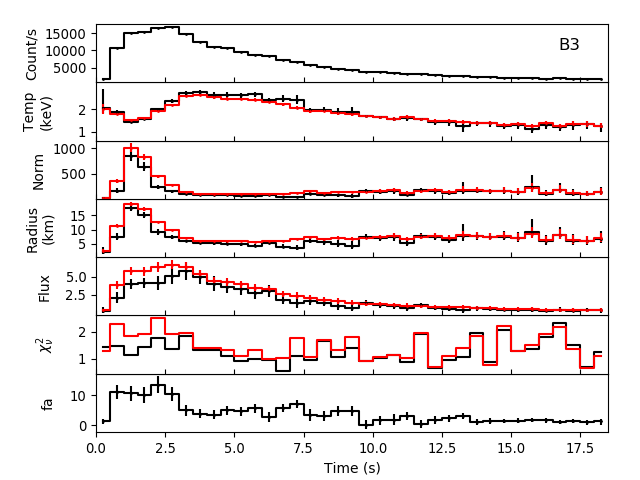}
\includegraphics[width=\columnwidth]{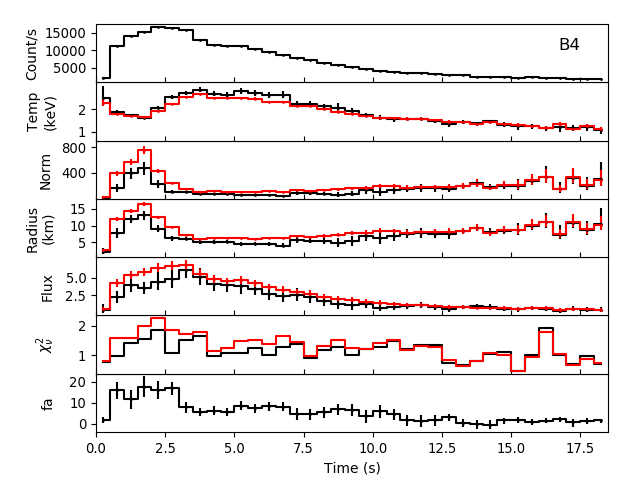}
\includegraphics[width=\columnwidth]{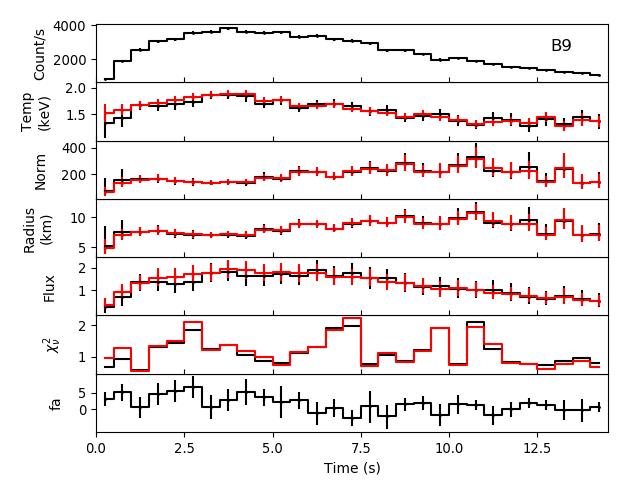}
\caption{Time-resolved burst profile in 3--25 keV band. Results from conventional method are shown in red, against the results from the $f_a$ method. Binsize is 0.5 second. Flux is specified in units of $10^{-8}$ erg~cm$^{-2}$~s$^{-1}$.~The three plots starting from top are for the three bursts B3,~B4 and B9,~respectively. The $f_a$ values indicate enhanced persistent emission near the burst peak. Refer \S\,\ref{sec-trs} for details.}
\label{TR-B3-B4-B9}
\end{figure}

Figure~\ref{TR-B3-B4-B9} shows the best-fit parameters obtained after performing time-resolved spectroscopy of 3 of the 12 bursts, which evince burst oscillations. The first panel of each plot shows count-rate during the X-ray burst, the second panel shows the temperature evolution, and the third panel shows the blackbody normalization during every 0.5~{\rm{s}} segment of the X-ray burst. The radius measured from the values of blackbody normalization is shown in the fourth panel. The unabsorbed bolometric flux (0.1--100~{\rm{keV}}), shown in the fifth panel of each plot, is given by (for each time step): $F_{\text{bol}}=\sigma T^4(R_{\text{bb}}/d_{10})^2=1.0763\times10^{-11}T_{\text{bb}}^4K_{\text{bb}}$~erg\,cm$^2$ s$^{-1}$ \citep{Galloway2006}. The sixth panel shows the reduced chi-squared ($\chi_{\nu}^2$) obtained from each spectral fitting. The temperature and the radius profile of bursts B3 and B4 suggest that these are PRE bursts. The error bars correspond to 90\%~confidence.\\

The values of maximum temperature noticed in the PRE bursts B3 and B4 are $2.64\pm0.05$~{\rm{keV}} and $2.68\pm0.05$~{\rm{keV}}, respectively. Other bursts show maximum temperature in the range 1.5--2.4~{\rm{keV}}. During the PRE phase, the photospheric radius is found to expand up to $19.1_{-0.7}^{+0.8}$ km in burst B3 and $16.5\pm0.6$ km in burst B4. Peak fluxes ($F_\text{max}$) are $6.71_{-0.65}^{+0.67}$ and $6.82_{-0.64}^{+0.66}$ $\times10^{-8}$ erg~cm$^{-2}$~s$^{-1}$ for PRE bursts B3 and B4 respectively. These flux values are consistent with the mean $F_\text{max}$ of $6.4\pm0.5$ $\times10^{-8}$ erg~cm$^{-2}$~s$^{-1}$ for PRE bursts observed by \rxte\ from \lmxb\ and using conventional method of spectral fitting \citep{Galloway2006}. For the non-PRE bursts, $F_\text{max}$ values lie between 0.6 and 3.6 $\times10^{-8}$ erg~cm$^{-2}$~s$^{-1}$. $\chi_{\nu}^2$ is $\sim$\,1.6--2.3 for the spectra in the PRE phase.\\

The peak bolometric flux during PRE can be used to estimate the source distance through $d=(L_\text{Edd}/4\pi F_\text{max})^{1/2}$ \citep{Barriere2015} where $L_\text{Edd}=3.79\pm0.15\times10^{38}$ erg~s$^{-1}$ for H-poor material i.e. $X=0$ \citep{Kuulkers2003}. We obtain a mean source distance of $6.8\pm0.4$ kpc. However, this $L_\text{Edd}$ value is valid only for very large photospheric radius expansion \citep{Galloway2008}. Following \cite{Galloway2020}, if we instead use $R=11.2 $ km in $1+z(R)$ in Equation 7 of \cite{Galloway2008}, we get $L_\text{Edd}=2.8\pm0.1\times10^{38}$ erg~s$^{-1}$ for $X=0$ and a source distance estimate of $5.9\pm0.3$ kpc.

\subsubsection{$f_a$ method}
\label{sec-trs-fa}
\cite{Worpel2013,Worpel2015} showed that the persistent emission evolves in the course of the burst as reflected in the dimensionless $f_a$ parameter. We apply the $f_a$ to both PRE and non-PRE bursts.~We use the \texttt{NTHCOMP} model \citep{Zdziarski1996,Zycki1999} with blackbody seed photons to model the 100 s preburst emission. \texttt{NTHCOMP} is a Comptonization model that is physical and convenient with a small number of parameters. It has two options: the seed photons can result from a single temperature blackbody (e.g., a boundary layer) or from a disk blackbody (e.g., an accretion disk). We choose the first option as it is found to provide a better fit than the second option. For the cases where the electron temperature is not well-constrained in this model, it is fixed to a value of 20~{\rm{keV}}.~The best fit values (see Table \ref{fa_bestfit}) are then fixed as required in the standard $f_a$ method and the blackbody model parameters are allowed to vary in the model: {\texttt{`TBABS*(BBODYRAD+CONSTANT*NTHCOMP)'}}. The value of the \texttt{CONSTANT} parameter here is regarded as the $f_a$ factor. Figure~\ref{TR-B3-B4-B9} also shows the results of spectral fitting using the $f_a$ method. The bolometric flux for the $f_a$ method is calculated as described in \S\,\ref{sec-trs-conv}. The evolution of the $f_a$ parameter is shown in the last panel.\\

The maximum temperature observed in a non-PRE burst using the $f_a$ method is $2.89^{+0.11}_{-0.10}$ in burst B2. The PRE bursts B3 and B4 show a maximum temperature of $2.75^{+0.09}_{-0.08}$ keV and $2.85^{+0.12}_{-0.11}$ keV, respectively. Other bursts show maximum temperature in the range of 1.5--2.5 keV. The maximum photospheric radius during the PRE phase is calculated to be $17.5_{-1.0}^{+1.1}$ km in burst B3 and $13.0\pm1.4$ km in burst B4. We find $F_\text{max}$ to be $5.79_{-1.15}^{+1.29}$ and $6.03_{-1.16}^{+1.29}$ $\times10^{-8}$ erg~cm$^{-2}$~s$^{-1}$ for PRE bursts B3 and B4, respectively. For the non-PRE bursts, $F_\text{max}$ values range from 0.6 to 3.5 $\times10^{-8}$ erg~cm$^{-2}$~s$^{-1}$.~A considerable overall improvement in the $\chi_{\nu}^2$ is found for the PRE bursts showing values in the range $\sim$\,1.1--1.8 while for the non-PRE bursts the $\chi_{\nu}^2$ values are comparable to that of the conventional method.\\

Using the peak flux obtained with this method, the source distance is estimated to be $7.3\pm0.8$ kpc using $L_\text{Edd}=3.79\pm0.15\times10^{38}$ erg~s$^{-1}$. For the alternate case of $L_\text{Edd}=2.8\pm0.1\times10^{38}$ erg~s$^{-1}$ mentioned earlier, the estimated distance is $6.3\pm0.7$ kpc.

\begin{table}
\caption{Fitting \texttt{NTHCOMP} parameters (abbreviated as `param.') for preburst emission of selected bursts (Energy: 3--25 keV; \texttt{TBABS}: $n_\text{H}=0.25\times10^{22}$ cm$^{-2}$). ``(f)'' indicates a ``frozen'' parameter. $kT_\text{e}$ and $kT_\text{seed}$ are stated in units of keV. Unabsorbed bolometric persistent flux (0.1--100 keV) is given in units of $10^{-9}$ erg~cm$^{-2}$~s$^{-1}$. For details, see \S\,\ref{sec-trs-fa}.}
\setlength{\tabcolsep}{2pt}
\renewcommand{\arraystretch}{1.4}
\begin{tabular}{>{\centering}p{0.15\columnwidth}|>{\centering}m{0.25\columnwidth}|>{\centering}m{0.25\columnwidth}|>{\centering\arraybackslash}m{0.25\columnwidth}}
\hline
Param. & B3 & B4 & B9\\
\hline
$\chi^2_\nu$ & 1.16 & 1.02 & 1.13\\
d.o.f. & 111 & 114 & 69\\
$F_\text{pers}$ & 2.89 & 2.41 & 1.43\\
\hline
$\Gamma_\text{nth}$ & $2.07_{-0.04}^{+0.04}$ & $1.85_{-0.04}^{+0.04}$ & $1.96_{-0.04}^{+0.04}$\\
$kT_\text{e}$ & $3.39_{-0.17}^{+0.19}$ & $2.55_{-0.09}^{+0.10}$ & 20.0 (f)\\
$kT_\text{seed}$ & 0.51 (f) & 0.25 (f) & 0.31 (f)\\
$K_\text{nth}$ & $0.22_{-0.01}^{+0.01}$ & $0.33_{-0.02}^{+0.03}$ & $0.11_{-0.01}^{+0.01}$\\
\hline
\end{tabular}
\label{fa_bestfit}
\end{table}

\section{Discussion and Conclusion}
\label{sec-dc}
In this work, we present spectral and timing analysis of 12 Type~I X-ray bursts in {\lmxb} using the data from {\lxp}.\\

Time-resolved burst spectroscopy is done with 0.5 s time-bin using both the conventional and the $f_a$ methods. The evolution of temperature and radius indicates a Photospheric Radius Expansion (PRE) during 2 of these bursts (B3 and B4). The ratio between peak and touchdown radius during PRE burst, the ratio of the temperatures corresponding to the peak radius and the touchdown radius, and similarly for the bolometric fluxes are noted in Table~\ref{peak-PRE}. Following the time $R_{\text{bb}}$ reaches its peak value, the flux continues to increase in both the PRE bursts. This was also noted in \cite{Sugimoto1984} for {\lmxb}~and subsequent studies with {\rxte}~\citep[e.g.,][]{Galloway2006}. The flux attains its maximum (flux peak) and starts to decrease prior to touchdown. The touchdown phase can be distinguished by high temperature (second maximum in the temperature profile) and a small emitting area. The ratio of $F_\text{bol}$ at peak radius and touchdown is lower with the $f_a$ method than that with conventional method because of different degree of enhancement of persistent flux (magnitude of $f_a$) at the two stages. The (radial) peak to touchdown delay is $\sim$\,2.5~s for PRE burst B3 and $\sim$\,2.0~s for PRE burst B4.~The values of peak fluxes and maximum radii for the PRE and the non-PRE bursts given in \S\,\ref{sec-trs} conform to those reported in \cite{Galloway2006} using {\rxte}~observations of \lmxb. Maximum flux, $F_\text{max}$, for both the PRE bursts, are higher than that for the non-PRE bursts.\\

From the spectral analysis using the $f_a$ method, we observe that the persistent emission is enhanced near the peak of the burst, especially for the PRE bursts in Obs~2 and Obs~3. The highest value of the $f_a$ parameter is seen in the PRE burst B4 to be $17.5\pm4.9$, whereas in the other PRE burst B3, it is $13.5\pm2.8$. For the non-PRE bursts, the highest $f_a$ value is $\lesssim$10. The $f_a$ values are consistent with those provided in \cite{Worpel2013}. The flux value in the $f_a$ method is lower than that in the conventional method around the peak of the burst since the calculated flux does not include the contribution of the scaled persistent emission to the total flux \citep{Worpel2013}. It should be noted that the maximum $f_a$ need not occur at the same moment as the peak radius \citep{Worpel2013}. We find that the two maxima concur in burst B4 but not in burst B3.\\

As has been observed earlier \citep[e.g.,][]{Bhattacharyya2018}, the $f_a$ method also gives a lower estimate for the radius as inferred from the blackbody normalization. The ratio of the touchdown radii as determined by the $f_a$ and conventional methods (i.e. $R_{\text{TD, }f_a}/R_\text{TD, conv.}$) is found to be $0.88_{-0.08}^{+0.09}$ and $0.84_{-0.08}^{+0.09}$ for PRE bursts B3 and B4, respectively. These values are comparable to the mean value of $0.97\pm0.11$ for the touchdown radii ratio obtained in \cite{Worpel2013} with PRE bursts from multiple sources. This difference in touchdown radii estimates poses more challenges to deriving neutron star parameters using radius expansion bursts.\\
 
We estimate the source distance using the maximum flux observed during the PRE bursts and obtain a value of $6.8\pm0.4$~\rm{kpc} with conventional method which is consistent with the known estimate of $6.0\pm0.5$~\rm{kpc} for this source \citep{Galloway2006}.~We also calculate the source distance using the assumption as in \cite{Galloway2020} and obtain an estimate of $5.9\pm0.3$~\rm{kpc} for the $X=0$ case,~in contrast to $5.0\pm0.5$~\rm{kpc} as quoted in their paper.\\

We detect burst oscillations in both the PRE bursts in the post-touchdown (PTD) phase, which corroborates previous findings \citep{Strohmayer1998,Zhang2013}. The oscillation frequency shows a $\sim$\,1 Hz drift which is consistent with $\lesssim$1\% frequency drift observed during Type~I X-ray bursts \citep{Strohmayer2001,Muno2002,Watts2012}. \cite{Zhang2013} found that for PRE bursts in \lmxb\, a PTD phase of length $\sim$\,2--8~{\rm{s}} (long PTD) is an indication of the presence of burst oscillations, where PTD length is defined as the contiguous time interval after the peak during which $R_\text{bb}$ remains more or less constant before increasing again. This trend was also noted for PRE bursts in 4U 1728\ensuremath{-}34 \citep{Zhang2016}. The PTD length is $\sim$\,4--5~{\rm{s}} for the PRE bursts in this study. Hence, the conjecture of \cite{Zhang2013} holds true in our analysis.\\

The frequency drift of 1~\rm{Hz} observed in our analysis can be explained with a simple model based on the conservation of angular momentum.~If a 10~\rm{m} thick layer of accumulated matter expands to about 30~\rm{m}, then the rotation frequency should decrease by $\delta{\nu}~{\approx}~ 2{\nu}(20~\text{m}/R)$,~where ${\nu}$ and $R$ are the stellar spin frequency and radius, respectively.~Therefore, in the beginning of these bursts we observe a burst oscillation frequency which is less than the stellar spin frequency and as the burning layer cools down, its thickness decreases and the burst oscillation frequency increases towards the stellar spin frequency in a few seconds \citep[for more details, see][]{Watts2012, Sudip2021}.~Similar frequency evolution (increase) of burst rise oscillations from \lmxb\ has also been observed earlier \citep{Sudip2005}.~However, this model has certain limitations specifically if the frequency drift is of the order of a few hertz \citep[see][for details]{Sudip2021}.\\

The oscillations in burst B9 are found to be particularly interesting.~A high rms amplitude of $34.0\pm6.7$\% is observed in the 3--25~\rm{keV} band during the initial 0.25 second of the rise. As shown in Figure~\ref{time-evol-ampl}, the amplitude shows a monotonic decrease until the burst peak, and later towards the declining phase of the burst at certain instances sudden rise in the fractional rms amplitude is observed.~The decrease of fractional amplitude with time during the rise and the concave shape of the fractional amplitude profile observed at 3${\sigma}$ confidence, provides by far the strongest evidence of thermonuclear flame spreading as observed with \emph{AstroSat}.~The concave shape of the fractional amplitude profile suggests latitude-dependent flame speeds, possibly due to the effects of the Coriolis force \citep[see, e.g.,][]{Strohmayer1997}.~A similar evidence of thermonuclear flame spreading on neutron stars has also been reported by \cite{Chakraborty2014} using the data from \emph{RXTE}--\textsc{PCA}.~\cite{Mahmoodifar2016} detected burst oscillations during both rise and decay portions of a burst, they found that the fractional rms amplitude shows a decreasing trend until the burst peak, then show an increase and finally decrease in the decay phase. Burst decay oscillations are often described using the surface mode model or the cooling wake model.~The fractional rms amplitude during the burst decay is typically small~($\sim$\,0.1) \citep{Bhattacharyya2010, Mahmoodifar2019} and lower values can be explained using the surface mode model. At the same time, it is not clear if this model can explain higher amplitude oscillations.~Therefore, it may be plausible that during the burst B9, both surface modes and cooling wakes are responsible for burst decay oscillations.\\

Apart from the hotspot, the cooler regions of the neutron star also contribute a low-energy persistent background flux to the oscillations, thereby decreasing their rms amplitude. The emission from the cooler regions is much less at higher energies, hence raising the rms amplitude of an oscillation.~Such a scenario can explain the observed correlation between amplitude and energy. Additionally, stellar rotation can also bring about such a correlation, albeit to a lesser degree through Doppler effect and phase lag between lower and higher energy pulses \citep{Muno2003}. The less dramatic amplitude versus energy slope such as those seen in the decay phase oscillations can be entirely a rotational effect. The observed phase lag of soft energy photons (soft lag) is seen as a result of the modulation of the emission from the hotspot by Doppler effect due to the rapid spin of the neutron star \citep[see, e.g.,][]{Muno2003,Chakraborty2017}.\\

On the other hand, burst oscillations in the PTD phase of PRE bursts could be due to the spreading of a cooling wake originating at higher latitudes of the neutron star so that the emission asymmetries take longer to dissipate \citep{Spitkovsky2002,Zhang2013}. The higher latitude origin is expected from high mass accretion rate \citep[see, e.g.,][]{Cooper2007} denoted by the high $S_a$ value of such PRE bursts in the color-color diagram (see \S\,\ref{sec-ccd}). The presence of the asymmetries on the rapidly rotating star, thus, gives rise to oscillations.~Cooling wake oscillations are expected to have smaller amplitudes than spreading hotspot oscillations which are observed during the rising phase \citep[see, e.g.,][]{Ootes2017}. This is reflected in our values of rms amplitudes.\\

Both the PRE bursts (B3 in Obs 2 and B4 in Obs 3) are seen during the soft state of the source indicated by their positions in the banana branch in Figure~\ref{color} at $S_a\sim2.1$. Burst oscillations are seen in both these bursts as well. For \lmxb, PRE bursts with oscillations preferably occur at $S_a\sim2.0-2.5$ \citep{Muno2004,Zhang2013}. One of the bursts (B9 in Obs 5) during the hard state at $S_a\sim1.3$ also shows strong $\sim$\,581 Hz oscillations.~Some other sources, beside \lmxb, which have been seen to evince burst oscillations in both hard and soft states are 4U 1702\ensuremath{-}429 (329 Hz), 4U 1728\ensuremath{-}34 (363 Hz), KS 1731\ensuremath{-}26 (524 Hz), Aql X--1 (549 Hz) and 4U 1608\ensuremath{-}52 (620 Hz), however, the majority of oscillations occurred during the soft state at $S_a>1.7$ \citep{Galloway2008,Ootes2017}. In the hotspot model perspective, this prevalence can be due to the ignition happening at higher latitudes for higher accretion rates and the effectiveness of Coriolis confinement of the flame off-equator, while an equatorial hotspot formed during lower accretion rates can get quickly wiped out \citep{Spitkovsky2002,Cavecchi2013,Cavecchi2015}.

\begin{table}
\caption{Comparison of the two stages in the two PRE bursts on the basis of time-resolved spectroscopy results. For details, see \S\,\ref{sec-dc}.}
\renewcommand{\arraystretch}{1.4}
\begin{tabular}{cccc}
\hline
Method & Peak/Touchdown & \multicolumn{2}{c}{Burst}\\
& ratio of & B3 & B4\\
\hline
\multirow{3}{*}{Conventional} & Radii & 3.13$_{-0.17}^{+0.18}$ & 2.74$_{-0.14}^{+0.15}$\\
& Temperatures & 0.58$_{-0.02}^{+0.02}$ & 0.61$_{-0.02}^{+0.02}$\\
& Bolometric Fluxes & 1.08$_{-0.16}^{+0.17}$ & 1.04$_{-0.15}^{+0.16}$\\
\hline
\multirow{4}{*}{Standard $f_a$} & Radii & 3.26$_{-0.32}^{+0.35}$ & 2.57$_{-0.35}^{+0.38}$\\
& Temperatures & 0.53$_{-0.02}^{+0.02}$ & 0.57$_{-0.03}^{+0.03}$\\
& Bolometric Fluxes & 0.82$_{-0.22}^{+0.23}$ & 0.70$_{-0.23}^{+0.24}$\\
& $f_a$ & $2.82_{-1.33}^{+1.29}$ & $3.10_{-1.46}^{+1.41}$\\
\hline
\end{tabular}
\label{peak-PRE}
\end{table}

\section{Summary}

\begin{itemize}

\item~\astrosat~observed~\lmxb~six times during its different spectral states. We analyze data of \textsc{LAXPC} onboard \astrosat\ and find 12 thermonuclear X-ray bursts in 6 observations made between July 2016 and August 2018.~Fast timing capability and large collecting area of \textsc{LAXPC} allow us to probe into the BO characteristics.~These oscillations are detected with 4--5${\sigma}$ confidence in three of the X-ray bursts.\\

\item The PRE bursts used in our study, show the presence of BOs during their decline phase while for the non-PRE burst, during both rise and decline phases oscillations are found. The initial 0.25 second of the burst rise show a very high value of rms amplitude ($34.0\pm6.7{\%}$).~The decreasing trend of the rms amplitude with time during the rise, observed at $3{\sigma}$ confidence, provides a strong observational support for flame spreading model.~Such a strong evidence has been found for the first time using \astrosat~data.\\

\item BOs in the PTD phase of PRE bursts could be due to the spreading of a cooling wake originating at higher latitudes of the neutron star.~The higher latitude origin is expected from high mass accretion rate as both the PRE bursts (B3 in Obs 2 and B4 in Obs 3) were seen during the soft state of the source indicated by their positions in the banana branch in Figure~\ref{color}.\\

\item Our time-resolved spectroscopy using the $f_a$ method, indicates enhanced contribution of the persistent emission near the peak of the burst, especially for the PRE bursts. The values of $f_a$ obtained are consistent 
with earlier reports \citep[see e.g.][]{Worpel2013}.~The touchdown radius  determined by the $f_a$ method is lower compared to the conventional method, indicating challenges in deriving neutron star parameters using radius expansion bursts.

\end{itemize}

\section*{Acknowledgements}

AB and PR acknowledge the financial support of ISRO under \emph{AstroSat} archival Data utilization program (No.DS-2B-13013(2)/4/2019-Sec. 2). This publication uses data from the \emph{AstroSat} mission of the Indian Space Research Organisation (ISRO), archived at the Indian Space Science Data Centre (ISSDC). We would like to thank Prof. H. M. Antia for discussions on the \textsc{LAXPC} data reduction software.~PR would also thank Rahul Sharma for discussions on the \textsc{LAXPC} data analysis. AB is grateful to the Royal Society, U.K. and to SERB (Science and Engineering Research Board), India, and is supported by an INSPIRE Faculty grant (DST/INSPIRE/04/2018/001265) by the Department of Science and Technology (DST), Govt. of India. We thank the referee for constructive comments and suggestions which have improved the paper.

\section*{Data Availability}

The data underlying this article are publicly available in ISSDC, at \text{https://astrobrowse.issdc.gov.in/astro\_archive/archive/Home.jsp}



\bibliographystyle{mnras}
\bibliography{sample} 





\appendix

\section{Additional Figures}


We present here phase-folded pulse profile during burst B9. We also give the Leahy-normalized power spectra with maximized signal power for bursts B3, B4 and B9.

\FloatBarrier

\begin{figure}
\centering
\includegraphics[width=0.95\columnwidth]{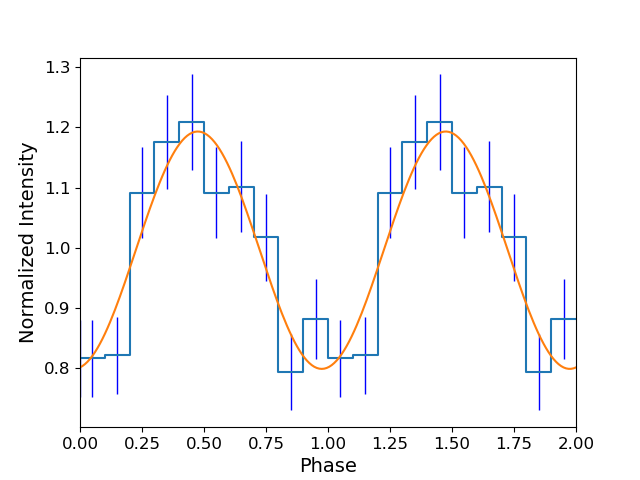}
\caption{Burst profile as a function of phase, plotted in the 3--25~{\rm{keV}} band for a 1~{\rm{s}} time window during the rising phase of burst B9. The smooth curve shows the sinusoidal fit with frequency 580.07~{\rm{Hz}}. The $\chi_{\nu}^2$ (d.o.f.) of the fit is 0.82 (17). The second cycle is shown for clarity. (see \S\,\ref{sec-entimebo})}
\label{A3-B9-BO-profile}
\end{figure}

\begin{figure}
\centering
\includegraphics[width=0.95\columnwidth]{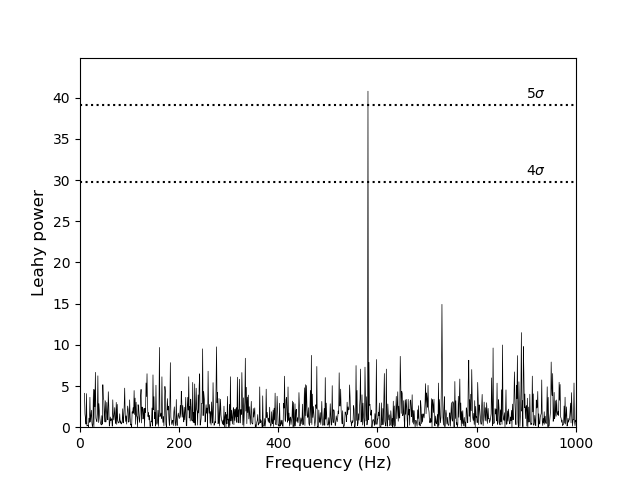}
\includegraphics[width=0.95\columnwidth]{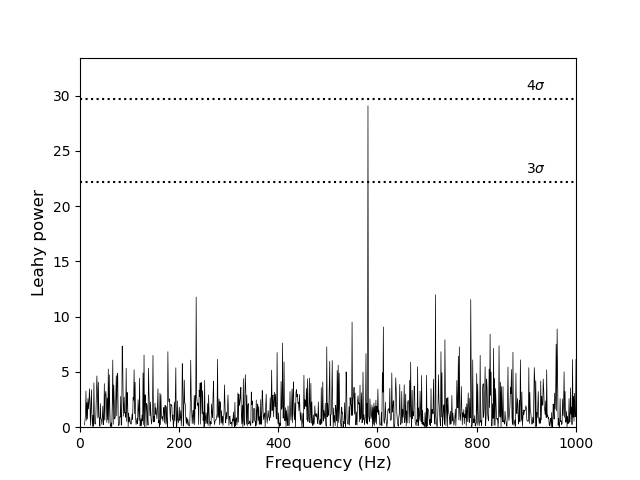}
\includegraphics[width=0.95\columnwidth]{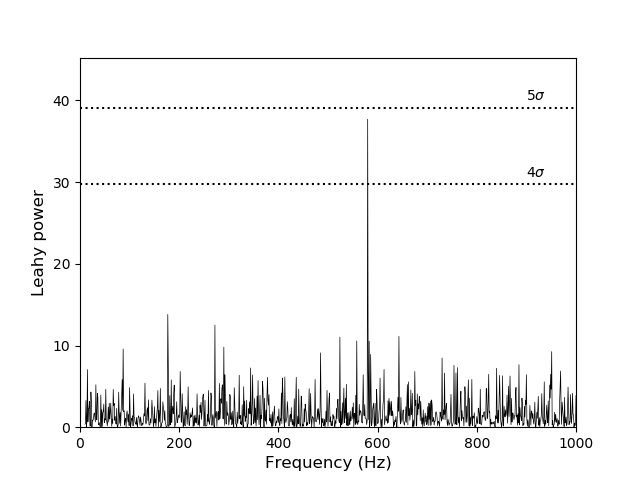}
\caption{\textit{Top Panel}: Power spectrum for a 2 s window of burst B3 in Obs 2 in the 8--25 keV band showing burst oscillations at 581 Hz. \textit{Middle Panel}:~Power spectrum for a 3 s window of burst B4 in Obs 3 in the 8--25 keV band showing burst oscillations at 581 Hz. \textit{Bottom Panel}: Power spectrum for a 1 s window during the rise phase of burst B9 in Obs 5 in the 3--25 keV band showing burst oscillations at 580 Hz. Refer \S\,\ref{sec-ps-bo} for details.}
\label{A2-B3-B4-B9-BO}
\end{figure}

\end{document}